\documentclass[conference,onecolumn]{IEEEtran}
\IEEEoverridecommandlockouts
\pdfoutput=1 
\usepackage{cite}
\usepackage{amsmath,amssymb,amsfonts}
\usepackage{algorithmic}
\usepackage{graphicx}
\usepackage{textcomp}
\usepackage{xcolor}
\usepackage{url}
\usepackage{hyperref}
\def\BibTeX{{\rm B\kern-.05em{\sc i\kern-.025em b}\kern-.08em
    T\kern-.1667em\lower.7ex\hbox{E}\kern-.125emX}}
\begin{document}

\title{Unlocking Metaverse-as-a-Service\\
The three pillars to watch: Privacy and Security, Edge Computing, and Blockchain\\}

\makeatletter
\newcommand{\linebreakand}{%
  \end{@IEEEauthorhalign}
  \hfill\mbox{}\par
  \mbox{}\hfill\begin{@IEEEauthorhalign}
}
\makeatother

\author{\IEEEauthorblockN{\textbf{Vesal Ahsani}}
\IEEEauthorblockA{Department of Electrical Engineering,\\ Sharif University of Technology, \\
Tehran, Iran \\
{\fontfamily{lmtt}\selectfont
vesal.ahsani@ee.sharif.edu}}
\and
\IEEEauthorblockN{\textbf{Ali Rahimi}}
\IEEEauthorblockA{Department of Electrical Engineering,\\ Sharif University of Technology, \\
Tehran, Iran \\
{\fontfamily{lmtt}\selectfont
a.rahimi@ee.sharif.edu}}
\linebreakand 
\IEEEauthorblockN{\textbf{Mehdi Letafati}}
\IEEEauthorblockA{Department of Electrical Engineering,\\ Sharif University of Technology, \\
Tehran, Iran \\
{\fontfamily{lmtt}\selectfont
mletafati@ee.sharif.edu}}
\and
\IEEEauthorblockN{\textbf{Babak Hossein Khalaj}}
\IEEEauthorblockA{Department of Electrical Engineering,\\ Sharif University of Technology, \\
Tehran, Iran \\
{\fontfamily{lmtt}\selectfont
khalaj@sharif.edu}}
}

\maketitle

\begin{abstract} 
In this article, the authors provide a comprehensive overview on three core pillars of metaverse-as-a-service (MaaS) platforms; privacy and security, edge computing, and blockchain technology. The article starts by investigating security aspects for the wireless access to the metaverse. Then it goes through the privacy and security issues inside the metaverse from data-centric, learning-centric, and human-centric points-of-view. The authors address private and secure mechanisms for privatizing sensitive data attributes and securing machine learning algorithms running in a distributed manner within the metaverse platforms. Novel visions and less-investigated methods are reviewed to help mobile network operators and metaverse service providers facilitate the realization of secure and private MaaS through different layers of the metaverse, ranging from the access layer to the social interactions among clients. Later in the article, it has been explained how the paradigm of edge computing can strengthen different aspects of the metaverse. Along with that, the challenges of using edge computing in the metaverse have been comprehensively investigated. Additionally, the paper has comprehensively investigated and analyzed 10 main challenges of MaaS platforms and thoroughly discussed how blockchain technology provides solutions for these constraints. At the final, future vision and directions, such as content-centric security and zero-trust metaverse, some blockchain’s unsolved challenges are also discussed to bring further insights for the network designers in the metaverse era.
\end{abstract}


\section{Introduction}
What is the \emph{metaverse}, exactly? The metaverse is a concept in the tech world that refers to a digital living environment where conventional social structures are changed. It is a term that combines the concepts of the Greek\footnote{\url{https://en.wikipedia.org/wiki/Meta}} prefix ``meta," which means ``more complete" or ``transcending," and the acronym ``Verse" for ``universe," which signifies a space-and-time container\footnote{\url{https://alldimensions.fandom.com/wiki/Category:Verse}}. The idea of the metaverse was introduced in Neal Stephenson's science fiction novel Snow Crash nearly 30 years ago \cite{joshua2017information}. The rapid advancements of technologies like blockchain, virtual and augmented reality, gaming, artificial intelligence, and the Internet of Things have made the metaverse one of the most buzzworthy terms in the tech world. Solutions and services are being developed for virtual worlds to allow users to have fun, intelligently engage with their surroundings, and form deeper connections with others \cite{Ahsani2}. Investment in the metaverse has grown significantly, with technology giants investing billions of dollars in its development and many businesses putting together their own plans for the metaverse. A McKinsey \& Company report predicts that the metaverse will be valued at over \$5 trillion by 2030 \cite{McKinsey_report}.

What is \emph{Metaverse-as-a-Service} (MaaS)? The phrase ``as-a-Service" originally appeared in a 1985 file with the United States Patent and Trademark Office (USPTO), and it gained popularity throughout the cloud computing era \cite{nguyen2012taxing}. Everything that may be considered as a service through a network can be referred to as XaaS \cite{banerjee2011everything}. Everything-as-a-service (XaaS) is a recent development in the information and communication technology sector that enables the provision of scalable computing resources on demand. Accordingly, the Metaverse can profit from ``as-a-service" models, in which the key elements and technologies of the Metaverse, such as platforms, infrastructures, software, and artificial intelligence (AI), could be provided as service models, i.e., Metaverse-as-a-service (MaaS). As tech giants entering the Metaverse space, including Microsoft, Samsung, NVIDIA, and others, it won't be long until there are many marketplaces with a Metaverse-as-a-Service (MaaS) offering allowing businesses to profit from the technology with lowered entry barriers. Even though MaaS is still a relatively new technology area, there are currently a number of vendors making headway there; such as Lovelace World, Propel MaaS, Touchcast, and MetaVerse Books. 

MaaS is characterized as an on-demand subscription solution that enables companies and/or operators to create and implement different forms (such as existence, management, coordination, and implementation) in the Metaverse to support the processing of Metaverse services, collaboration, company operations, and products, among other related scenarios. Everything in Metaverse can be thought of as a delivery model that can be simply generated and/or modified as function modules, similar to XaaS in cloud computing systems. See Figure \ref{three_pillars}.

\emph{The main benefits of using MaaS can be summarized as follows}:
\begin{itemize}
\item Products for the Metaverse may be created by businesses without substantial digital experience or knowledge. Without prohibitive capital requirements, even small to mid-sized firms can engage in the Metaverse economy.
\item It promotes financial investment in a still-developing technology. The majority of systems are currently only designed for consumer usage, and solutions like Microsoft Mesh and the Horizons app suite from Meta have not yet been widely released. In this setting, MaaS enables businesses to make low-risk investments and profit from technology. This model also reduces the time of programming, setting up and installing systems and brings the investor a profit sooner.
\item MaaS may ultimately lead to industry standardization, with selected few businesses serving as Metaverse ``brokers" to aid in infrastructure creation.
\item MaaS model is a win-win situation. The seller of this service does research, programming and implementation only once, but can sell this service many times. On the other hand, the buyer of the service can also use Metaverse at a competitive cost without getting into the technical, management and implementation complexities of Metaverse. 
\end{itemize}

\begin{figure}[h]
\includegraphics[width=\textwidth]{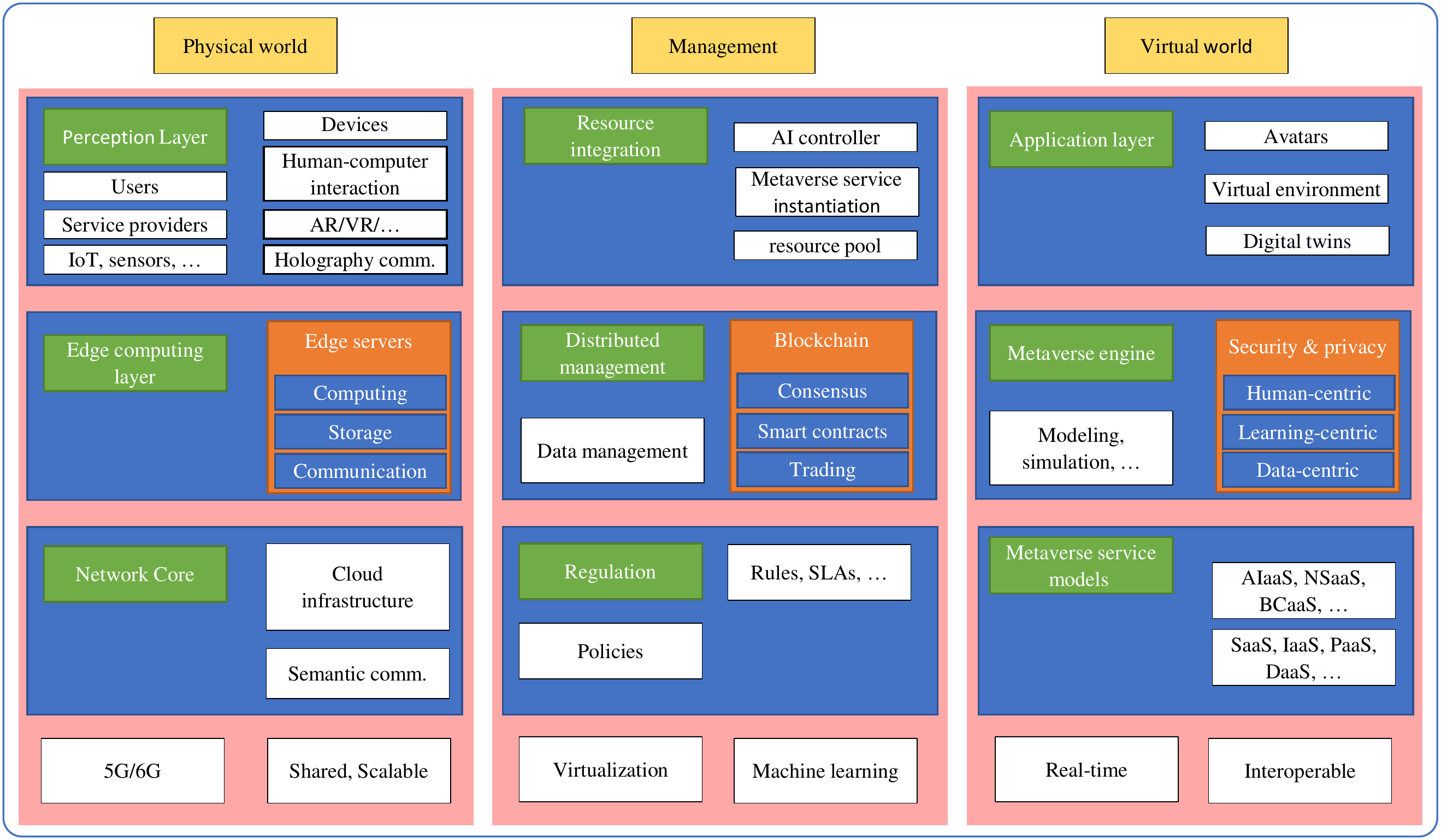}
\caption{Structure and important modules in the Metaverse network.\label{three_pillars}}
\end{figure}

Despite the outstanding advances in today’s existing MaaS platforms, the Metaverse would still need a more comprehensive and robust collection of standards and protocols to embrace interoperability, much more comprehensively than what the current internet includes, based on a set of rules and policies for communication, visuals, graphical demonstration, and data. For instance, Fortnite runs on practically all popular platforms (such as iOS, Android, PlayStation, and Xbox) and supports numerous identity/account systems and payment options, which forces rivals to cooperate (i.e., engage in interoperability). Web2 goliaths like Apple and Google employ similar technology today, but they are not built to integrate with one another. Building a scalable Metaverse will depend more on interoperability than anything else. Companies might create their own Metaverse campuses using established protocols with the support of Metaverse-as-a-Service (MaaS), and then start providing immersive experiences that promote social interaction. The metaverse infrastructure should be supported by extensive accessibility to mediate various aspects of human-beings’ lives, to be able to provide immersive experiences. This opens up new venues for privacy and security risks. 

The Internet-of-Things (IoT) plays an important role in digitalizing the physical world by utilizing pervasive sensors, cameras, wearables, etc. Networking connectivity is then provided via wireless networks, while the computing and storage are provisioned through cloud and edge computing. The IoT networks act as ``bridges'' between the physical world and its digital counterpart \cite{hu2020cellular}. The information flow between the two worlds, can help facilitate the decision making in both physical and digital worlds. Users, who are mainly represented as avatars, can also produce and exchange digital contents across various platforms in the metaverse. Mobile users interact with the digital world via their smartphones, wearable devices, and augmented reality/virtual reality (AR/VR) helmets, to create and share contents and gain knowledge.

Information is the core resource of the metaverse. The data flow within the employed networks of metaverse has the key role in realizing the integration of physical and digital worlds. To better understand the security and privacy needs in the metaverse, we indicate that there exist two main sources of information in this era: The first one is the data gathered from and exchanged by the real world that might be utilized and visualized digitally in the virtual space. The second source of information is the output of the virtual worlds, e.g., the information generated by digital assets and services in a MaaS platform. At the same time, artificial intelligence and machine learning (AI/ML) algorithms, performed in the computation layer or the digital twin layer, help facilitate rendering and offering various services. Accordingly, it is crucial to safeguard the privacy and security of data flows within the MaaS platforms, as well as the learning algorithms. 

On the other hand, performing all metaverse computations on cloud servers is not necessarily the best option. In some situations, using edge servers close to users can facilitate low-latency services, reduce network traffic, help manage data, and improve user experience. In addition, the paradigm of edge computing is closely related to other technologies such as artificial intelligence and IoT. For example, with the help of edge servers, it is possible to train neural networks locally with the participation of end devices without having any internet connection. Also, edge servers enable secure management and access control of IoT devices and sensors on-premise.

Last but not least, blockchain is regarded as one of the metaverse's core infrastructures and helps supply the metaverse with laws that are clear, open, effective, and trustworthy \cite{yang2022fusing} because it can connect disparate minor sectors and create a solid economic system,. It will be challenging to determine the worth of the commodities and resources traded in the metaverse without the assistance of blockchain technology, especially when those digital components interact economically with the real-world economy. Therefore, it would be wise to investigate blockchain technology for MaaS platforms along with the other two pillars; privacy and security and edge computing.

In the following, we provide a comprehensive review on guidelines to safeguard the privacy and security of MaaS platforms from different perspectives. Additionally, in order to help metaverse operators to identify an appropriate approach for using edge computing in the metaverse, we have focused on the advantages and challenges of using the edge computing paradigm in MaaS in two separate subsections. Moreover, we thoroughly discuss the requirments of using blockchain technology and the actions MaaS developers should take to solve many technical challeneges by using blockcahin. Finally, we conclude the paper with future visions and directions.

\section{Privacy and Security}
\subsection{An Overview of Privacy and Security Challenges for the metaverse}
Despite the ever-increasing advances in developing numerous services for the metaverse, privacy and security challenges are considered as the main concerns that need to be properly understood. Considering the fact that realizing the concept of MaaS requires an extremely wide variety of computation, communication, and networking, a wide range of security and privacy threats also arise in the metaverse. See Figure \ref{PS_figure}.

\begin{figure}[h]
\includegraphics[width=\textwidth]{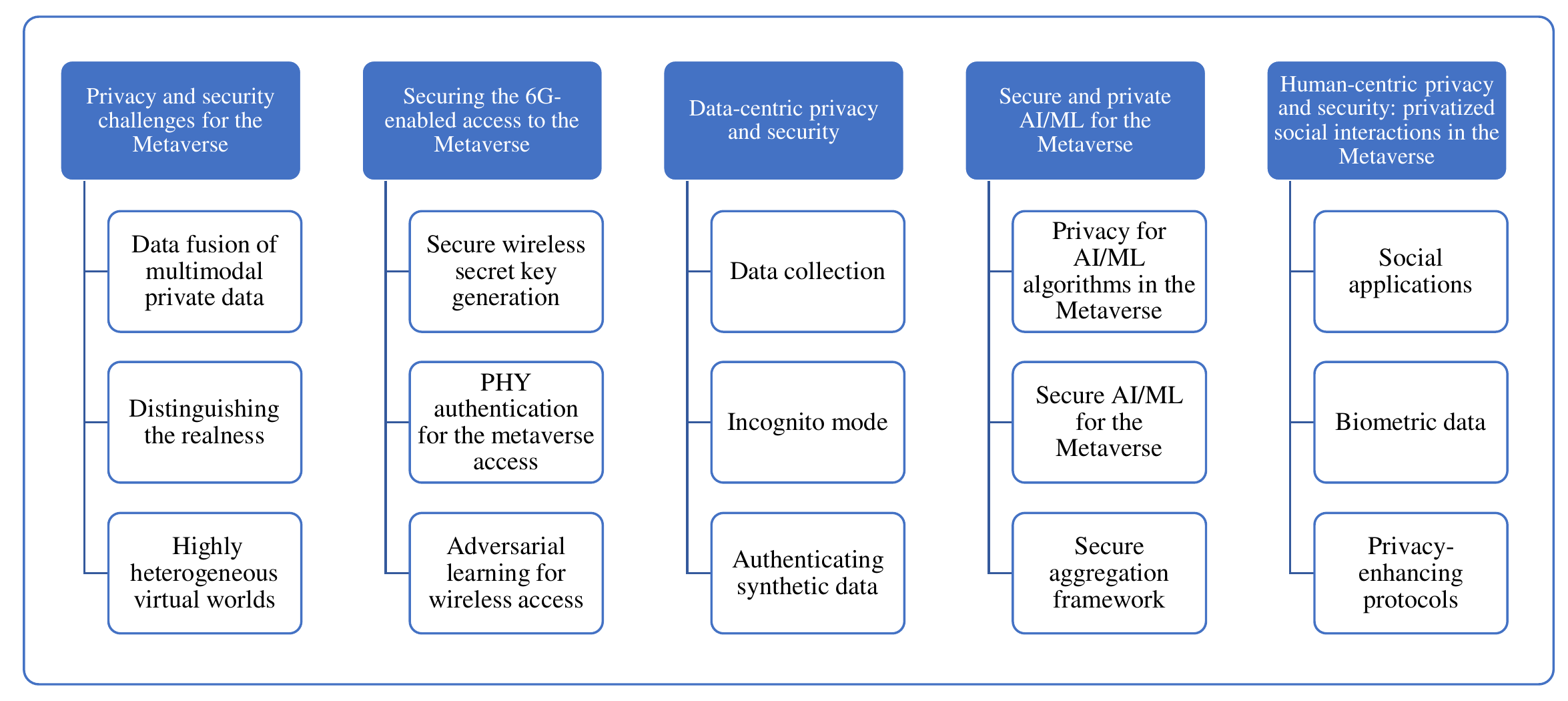}
\caption{Privacy and security for the Metaverse.\label{PS_figure}}
\end{figure}

A variety of recent technologies are integrated into the metaverse as its basis, hence, the
intrinsic vulnerabilities of them may be inherited by the corresponding MaaS platforms. Recently, numerous privacy and security flaws have been identified for the emerging technologies, including the vulnerabilities of cryptography-based key management schemes against quantum computers \cite{porambage2021roadmap, chorti2022context
}, the privacy leakage of distributed learning networks against honest-but-curious servers or malicious clients \cite{kadhe2020fastsecagg}, and the misuse of massive private data by service providers \cite{falchuk2018social}, just to name a few. Such threats can be intensified in the virtual world, while simultaneously other threats can also happen, which does not exist in the physical world, e.g., virtual spying \cite{falchuk2018social, Washingtonpost}. Notably, massive amount of sensitive data are utilized to create a digital replica of the real world. This opens up new challenges in terms of threats to privacy and security. Individuals use wearables or AR/VR devices with built-in sensors collecting biometric data, brain patterns, users’ speech and facial expressions and poses, and also the surrounding environment in order to render a high-quality immersive experience for the MaaS users. In addition, clients are under the threat of being illegally tracked through their VR headsets or wearables. Such a miscellaneous set of sensitive information flow in the context of metaverse services provides adversaries with a broad attacking surface which requires proposing novel secrecy- and privacy-preserving mechanisms for mobile network operators and service providers.

The main privacy and security challenges in the era of metaverse can be identified as follows.
\begin{itemize}
\item Data fusion of multimodal private data corresponding to interactions between users, avatars, and environments imposes serious challenges for the MaaS platforms. Collection, communication, and processing of such sensitive information must be taken into consideration in terms of privacy and security risks.
\item The blurring boundary between the real world and its virtual counterpart causes serious confusions in terms of distinguishing the realness. Therefore, we should focus on providing stringent authenticity guarantees from the MaaS providers’ perspective.
\item Virtual worlds in the metaverse are highly heterogeneous in terms of hardware implementation, communication, and learning interfaces. The beyond-the-fifth and the sixth generation (B5G/6G) of wireless communication networks, as the main fabric to facilitate the communications within the metaverse, pose critical challenges for conventional security solutions. This raises urgent needs for proposing context-aware and flexible security and privacy mechanisms \cite{chorti2022context, letafati2021learning}.
\end{itemize}
We aim to carefully review these items in this paper. We provide readers with useful insights through a comprehensive overview of security solutions across different layers of the MaaS platforms, ranging from the access layer to the social aspects. We also review different solutions in the contexts of communication, computation, and learning.

\subsection{Securing the 6G-enabled Access to the Metaverse}
For the deployment of access layer to the metaverse, the sixth generation (6G) of wireless
networking will play a pivotal role. 6G technologies, such as pencil-sharp beams, edge AI, and integrated sensing and communications (ISAC), provides users with perceived understanding of their surroundings. These technologies are the key enablers for implementing novel PHY layer security (PLS) schemes as promising approaches to safeguard the security of the metaverse access, while offering information-theoretic security guarantees \cite{letafati2020lightweight}. Leveraging PLS, intelligent, flexible, and context-aware mechanisms are feasible for detection and mitigation of privacy and security vulnerabilities \cite{chorti2022context, letafati2021learning, letafati2021deep, letafati2022hardware}. Considering the new era of truly end-to-end (E2E) quality-of-experience provisioning for the MaaS products, service level agreements (SLAs) are expected to include guarantees about the quality of security (QoSec) as well \cite{letafati2021learning}. This will include addressing the required security level, adaptive, risk-aware, and adjustable security solutions, just to name a few \cite{chorti2022context}. The evolution of the metaverse systems is expected to introduce new means to harvest and interpret the ``context'' of the communication, where PLS techniques can be considered as a promising QoSec-assuring approach.

In order to secure the metaverse access, we introduce the following security mechanisms,
which can provide lightweight and scalable solutions for securing the access to the MaaS
platforms.

\subsubsection{Wireless Secret Key Generation and the Role of PHY Security} 
Smart wearable devices and cameras, together with head-mounted displays, e.g., Oculus helmet and Vive Pro headsets, and handheld controllers are considered as the main terminals for entering the metaverse. In this regard, key generation and management is essentially required for smart devices to establish secure connections for accessing the metaverse, transmit sensory data, and receive immersive experiences. In \cite{sun2020accelerometer} and \cite{chen2018lirek}, intrinsic features of specific wearable devices, such as gestures and motions, are taken into account to propose key agreement schemes. Researchers in \cite{zhang2021h2k} propose a heartbeat-based key generation scheme based on the measurements from electrocardiography (ECG) and photoplethysmography (PPG) sensors as the source of common randomness.

While the security mechanisms of the 5G standard rely on cryptography-based keys, such as the elliptic curve cryptography (ECC) to fulfill the confidentiality and authentication requirements, 6G networks with many peer-to-peer communications undermine the performance of conventional solutions. Besides, assumptions on the security of cryptographic methods based on the mathematical hardness of solving a certain problem will no longer hold with the advent of quantum computers. To efficiently secure the metaverse era, data communications should to be proactively secured, where PLS has such capabilities for quantum-resilient security \cite{chorti2022context}. As a promising framework to migrate from the conventional complexity-based solutions towards lightweight security techniques, PHY layer secret key generation (PHY-SKG) has been envisioned to be employed for 6G networks \cite{letafati2021deep, letafati2022hardware}. PHY-SKG is realized through implementing lightweight mechanisms with minimum required changes at the control plane. This can actually provide the access networks with different advantages. In particular, PHY-SKG is thoroughly decentralized without relying on any infrastructure of a particular entity, which can substantially reduce the required time for key agreement, making it suitable for extremely low latency applications in the metaverse. Moreover, continuous update of the key is realizable thanks to the dynamic variations of the PHY attributes. The PHY-SKG protocol under realistic assumptions of hardware impairments and adversarial attacks is developed in \cite{letafati2021deep}. Notably, the PHY-SKG has shown to be resilient against man-in-the-middle adversarial attacks in \cite{letafati2022hardware}. Furthermore, for establishing secure communication between digital healthcare devices and the access point, a lightweight learning-based key agreement protocol is proposed by the authors in \cite{letafati2022wireless}, which can guide designers to bring device-level intelligence for securing the metaverse access.

\subsubsection{Adversarial Learning for Wireless Access}
Due to the proliferation of virtualized services in the metaverse era, malicious actors may have access to the network functions, such as the open-source software of virtualized access networks, and can exploit vulnerabilities of the metaverse access through adversarial machine learning (AML) \cite{geiping2021doesn}. Adversaries can poison the wireless access of the metaverse, by manipulating the inputs to the learning algorithms employed for authorizing users \cite{shi2021adversarial}, leading to unauthorized access to the metaverse network due to the mislead access model. Malicious users can also infer sensitive information about edge devices, users, and applications which have access to the network, through membership attribute inference attacks \cite{shi2022membership} and model inversion attacks \cite{geiping2020inverting}. As a privacy threat to the MaaS access, data characteristics such as device-level information may leak to adversaries. Malicious users can exploit this leaked information using membership inference attacks \cite{shi2022membership} by building an inference model to determine whether a sample of interest (associated with a particular device) has been used in the training data of the MaaS provider. To mitigate such attacks, generative adversarial networks (GANs) are shown to be capable of detecting anomalies and mitigating wireless attacks for the next generation of communication and networking services \cite{ayanoglu2022machine}. 

Connection requests for the metaverse services come with quality of experience (QoE) requirements such as throughput and latency, which can be fulfilled by assigning radio resource blocks and processing power to the requests. Although network slicing can manage the MaaS QoE requirements, AML techniques have the potential to attack network resource management and disrupt B5G network slicing, which can impose huge losses for the network operators, if not properly secured \cite{shi2021adversarial}. To combat such attacks, different reactive and proactive defense mechanisms are proposed in \cite{shi2021attack}, including: the induction of randomness to the decision processes to reinforce robustness against malicious nodes.

So far, the security challenges for accessing the metaverse were addressed. In the following sections, we go deeper into the privacy and security aspects inside the metaverse, in a data-centric manner. We also address synthetic and authentic realness, envisioning the privacy enhancement within the metaverse platforms.

\subsection{Data-centric Privacy and Security}
Inside the metaverse, the scope of data collection and processing far exceed what is traditionally performed in mobile and web applications. The technologies employed for the metaverse do not just track where users click, but where they go, what they do, whom they interact with, what they look at, etc. Therefore, it is crucial to the users’ privacy and security to implement data-centric privacy-preserving mechanisms and push for meaningful, sensible, and aggressive regulations in terms of data privacy.

Recent studies demonstrate how easily the metaverse users’ privacy can be compromised \cite{nair2022exploring}. It is shown that an adversarial program can accurately infer over 25 personal data attributes. To deal with the unprecedented data-centric privacy risks in the metaverse platforms, the idea of ``incognito mode'' has recently been developed for the metaverse as a novel and promising solution for safeguarding VR users’ privacy \cite{nair2022going}. To realize the incognito mode in the metaverse, client-level differential privacy framework is proposed which has been shown to be capable of protecting a wide variety of sensitive data attributes in VR applications. It is show in \cite{nair2022going} that by employing randomized mechanisms (depending on the format of the private attributes), sensitive user data attributes can be obscured effectively \cite{nair2022going}.

We are facing a new trend of intelligent systems that are empowered by synthetic realness, in which AI-generated data is vastly exploited to reflect various aspects of the physical world \cite{Letafati28, solaiyappan2022machine}. Inside the metaverse, consumers are not simply targeted via pop-up ads, rather, they are provided with ``immersive contents'' in the form of virtual people and activities that look realistic \cite{nightingale2022ai}. Nevertheless, we should concern about the authenticity of synthetic data at different levels. Authenticating synthetic data can make AI algorithms more fair and secure through correcting data biases and safeguarding data privacy \cite{Letafati30}. Synthetic data is utilized for training AI models in ways that real-world data practically cannot or should not, while protecting privacy and confidentiality. It is critically important for the network operators and MetaaS providers to leverage generative AI in an authentic way to maintain trust for their customers \cite{Letafati32}. To realize authentic realness, the provenance, the policy, and the purpose of utilizing synthetic data should be taken into account by companies and service providers \cite{rosenberg2022regulating}. To this end, distributed ledger technology (DLT) can be employed to verify the provenance of digital content, hence addressing the authenticity. Moreover, network operators should clarify the purpose behind the exploitation of synthetic content and the resulting advantage over non-synthetic content.

\subsection{Secure and Private AI/ML for the Metaverse}
In this section, we focus on challenges and solutions for the privacy and security of AI/ML
implementations in the metaverse.

Although metaverse can bring amazing AI/ML-based services to individuals and enterprises, the utilized learning algorithms are still vulnerable to privacy and security risks. There exist challenging bottlenecks for efficient deployment of AI/ML techniques. Among these challenges, the big concern is that the learning algorithms might leak users’ private data \cite{wang2020attack}. Besides, users might not be willing to share their private data in the metaverses, making it challenging for AI/ML algorithms to perform a comprehensive data analysis for enhancing the learning-based services of the metaverse platforms. To address the privacy risks of AI/ML algorithms, the integration of federated learning (FL) \cite{li2020federated} and cross-chain technologies can be considered as a promising solution to design and implement privacy-aware frameworks for the MaaS platforms \cite{jiang2021cooperative, kang2022blockchain, lu2020low}. For instance, \cite{lu2020low} proposes a hierarchical blockchain-based framework for decentralized FL. A main chain is employed to mimic the parameter server (aggregator server) of the FL algorithm, and multiple subchains are implemented to manage local model updates generated by smart devices (or their digital counterpart) acting as workers.

Despite the fact that FL algorithms leverage locally-trained models instead of the raw data of metaverse participants, sensitive information can still be inferred by analyzing the model parameters uploaded by clients \cite{geiping2020inverting}. If the model updates are inspected by bad actors, participant users’ privacy and security would be threatened. In addition, there might exist some malicious users, who proactively upload adversarial data to the aggregator server(s) to mislead the training process, known as backdoor attacks \cite{wang2020attack}. FL still has limitations in terms of security and privacy: First, malicious clients may inject mislabeled or manipulated data to mislead the global model, which is known as data poisoning attacks \cite{geiping2021doesn, lyu2020threats}; Second, during model update of AI/ML algorithms, adversarial participants can save the model parameters and infer sensitive information, e.g., private attributes of participants’ data. They might also be able to recover the original training samples of users \cite{geiping2020inverting, lu2020low}. In the following, we discuss the possible countermeasures to the abovementioned threats in more details.

We take into account the fundamental fact that local model updates in
distributed AI/ML algorithms carry extensive information about the local datasets owned by the metaverse clients. Hence, secure aggregation of local models are necessary for the distributed learning entities in the metaverse. Moreover, there might exist some adversarial participants in the metaverse platform who actively upload “poisoned data” to the servers (as aggregator entities) to mislead the training processes. Such attacking venues can provide the surface for the so called “backdoor attacks” on the collaborative model training procedures within the MaaS platforms \cite{wang2020attack}. As an example, virtual service providers and mobile network operators employ learning algorithms as a service for their users \cite{Letafati43}. Malicious users have the capability to eavesdrop on the models exchanged to infer the membership of specific clients or undermine the performance of AI/ML services \cite{shi2022membership}. In addition, attackers can hack edge devices in the physical world or impersonate benign avatars as a man-in-the-middle. They can then inject adversarial training sample to spoof the learning services within the MaaS products.

In the following, we address secure aggregation framework to protect learning tasks from
being eavesdropped by malicious participants.

\textbf{Secure Model Aggregation:} Generally speaking, state-of-the-art secure aggregation (SA) protocols rely on two main principles: i) pairwise random-seed agreement between clients to generate “masks” for hiding the metaverse users’ models; and ii) secret sharing of the random seeds. A SA protocol enables secure computation of distributed learning models, while ensuring that the aggregator entities do not infer information about the local models \cite{so2022lightsecagg}. Taking into account the fact that the MaaS users might be “honest-but-curious” during the learning process, a SA protocol must guarantee that nothing can be learned beyond the aggregated learning model, even if up to $T$ users cooperate with each other. This is called $T$-privacy property of SA algorithms. In \cite{so2022lightsecagg}, a modular system design for SA is proposed, which can help develop lightweight SA protocols for the learning infrastructure of the MaaS platforms.

\textbf{Backdoor Attacks:} Malicious clients may inject mislabeled or manipulated data to mislead the learning models within the highly-distributed and heterogeneous infrastructure of the metaverse platforms, which is known as data poisoning attacks \cite{geiping2021doesn, lyu2020threats}. In this context, backdoor attack is a type of “data poisoning” attacks that aims to tamper a subset of training data via injecting adversarial triggers, such that AI/ML models trained on the manipulated dataset make (targeted) incorrect prediction/decisions during the inference \cite{lyu2020threats}.

Backdoor attack is aimed to mislead the trained model to infer a target label on any input data that has an adversarially-chosen embedded pattern (a.k.a trigger). Distributed version of backdoor attacks, known as DBA, decomposes a global trigger pattern into separate local patterns and “embed” them into the training set of different adversarial parties respectively. This is in contrast to the conventional attacks on the surface of distributed AI/ML algorithms \cite{bagdasaryan2020backdoor}, where a unified global trigger pattern is embedded for all malicious parties. The experiments in \cite{xie2019dba} show that the DBA method outperforms centralized attack significantly when evaluated with the global trigger.

To overcome backdoor attacks, adversarial training (AT) is currently considered as an empirically strong defense mechanism \cite{lyu2020threats}. AT corresponds to the mechanism of training a model on adversarial examples, with the aim of making the learning service more robust to attacks and reducing the inference error on benign samples. AT is comprehensively reviewed in a wide variety of researches, including \cite{lyu2020threats, goodfellow2018making, kurakin2016adversarial}. Recently, a generic framework to defend against the general type of data poisoning attacks is proposed in \cite{lyu2020threats} by desensitizing networks to the effects of poisoning attacks. To mention an application of AT within the metaverse, an AT-based protection against facial recognition systems is proposed in \cite{cherepanova2021lowkey}, which can be effectively utilized for the MaaS platforms.

\subsection{Human-centric Privacy and Security:‌ Privatized Social Interactions in the Metaverse}
In the metaverse, human users socially interact with others through their avatars and experience MaaS platforms by leveraging ubiquitous smart devices and network access in a real-time manner \cite{duan2021metaverse}. The metaverse as a “human-centric” virtual environment is supported by massive human-centric social applications, such as holography AR, immersive VR gaming, and the tactile Internet. Such immersive services can be realized through the interdisciplinary communication-computation-storage co-design techniques, integrated with the concepts of human perception, cognition, and physiology for haptic communication \cite{antonakoglou2018toward}. Hence, the most sensitive and personalized information will be exchanged among different parties, which significantly highlights the essence of employing novel intelligent privacy- and secrecy-preserving algorithms for the information security and social privacy of MaaS users \cite{buck2021privacy}.

The biometric data plays a significant role in terms of social interactions inside the metaverse. To elaborate, human-driven agents and avatars, with whom the users interact, are being developed based on users’ personal data \cite{buck2022security}. Such sensitive data can be fed into AI/ML algorithms to create personalized “interaction partners” and influence social interaction behaviors of the metaverse users. By exploiting physical and mental traits inferred from biometric data, the interaction partners can maliciously lead users into undesired situations and behaviors. Hence, the metaverse users can be manipulated not only by businesses and institutions, but by other individuals in a human-centric manner.

To mitigate such human-centric threats, attention should be paid to the users themselves. Metaverse service providers should make users aware of the implications of biometric data extracted by AR/VR devices. Developers can implement privacy-enhancing protocols such as differentially private mechanisms as proposed in \cite{wei2020ldp} and \cite{schmidt2022mitigating}. Users can also be provided with secondary avatars, e.g. clone, to hide their actions in the metaverse \cite{falchuk2018social}. The secondary avatars help users hide their real behavior within the metaverse. Furthermore, users must be able to have an adjustable level of privacy \cite{nair2022going} to dynamically manage their personal space in the virtual world by choosing their desired privacy configuration. From a governance layer, a modular bottom-up governance approach is proposed in \cite{schneider2021modular}, which can be adapted to different platforms and use cases. Decentralized autonomous organizations (DAOs) are examples of governance systems that allow users to be involved in decision-making processes for the MaaS platforms \cite{Letafati56}. In this regard, organizations such as XR Safety Initiative (XRSI)\footnote{https://xrsi.org} will play an important role in encouraging MaaS providers and network operators to help facilitate enhancing trust within the MaaS implementations in terms of human-centric privacy and regulations.

\section{Edge Computing}

To unlock the full potential of the metaverse, edge computing represents a promising paradigm for technical development of the metaverse. Edge computing is a well-known technology that allows operators to perform processing, storage, and network functions close to the edge of the network, where users are located. By using this paradigm in the Metaverse, network efficiency is improved, costs are reduced, and new features are added.

As an illustrative example, consider a smart hospital in the Metaverse. In this center, telemedicine facilities are implemented. Artificial intelligence is used to diagnose diseases and suggest treatment, hospitalized patients are monitored in real time with IoT sensors, and remote surgery can be done in this center. In this scenario, if all this data is to be sent directly to the cloud server for processing and storage in a traditional way significant network bandwidth will be occupied. In addition, with this method, there will be a long delay in sending data, and it is possible that the confidentiality of patients' data will be compromised. This motivates the need for the edge computing paradigm that can process and store data near the edge of the network where data is generated. In short, this design can be such that several servers are installed in the hospital that can quickly collect and process large amounts of data, feed it to neural network models, then send the extracted information to the cloud server.

Edge computing can be a lucrative opportunity for network operators. One-third of network operators' costs are spent on real estate to house infrastructure \cite{sabella2019multi}. Converting these locations into edge computing sites is a huge revenue potential for network operators. They can also use their extra processing and storage resources to empower network edge fascilities. On the other hand, operators can have a successful entry in the development of metaverse. They can play key roles by using the services they provide in the 5G and 6G mobile networks \cite{operators_seven}. Consequently, providing MaaS is a smart way to boost operator income and develop diverse applications. In the MaaS model, organizations can flexibly create their own virtual world according to their needs and with the most recent versions of the tools. This is made possible by using operator edge facilities and networks. In addition to the network operator that provides a lot of hardware and software infrastructure, vendors such as application developers, edge computing equipment vendors, system integrators, and edge computing solution providers are also active in this market and new revenue lines are open to them. Furthermore, various vertical markets such as automotive vehicles, virtual content productions, robotics, eHealth, entertainment, smart manufacturing, smart grids, and many others are added to this market. For a survey on the edge computing market, see \cite{Edgemarket}.

An edge computing system includes the internet gateway, edge servers, and perception layer \cite{kumari2022edge}. Internet gateway is responsible for connecting the edge system with the network backbone and cloud servers. Edge servers process, store, and aggregate data. Perception layer includes edge devices such as IoT sensors and actuators, AR glasses, VR headsets, haptic gloves, smartphones, wearable devices, personal computers, and holographic displays. The metaverse can benefit from edge computing in different ways. See Figure \ref{Edge_figure}. In the following, we explain some of them. In addition, edge computing also brings challenges. After explaining the advantages, we explain the challenges of using edge computing in the metaverse. It should be noted that we provide a high-level review of these items and delving into their technical details is beyond the scope of this paper.

\begin{figure}[h]
\includegraphics[width=\textwidth]{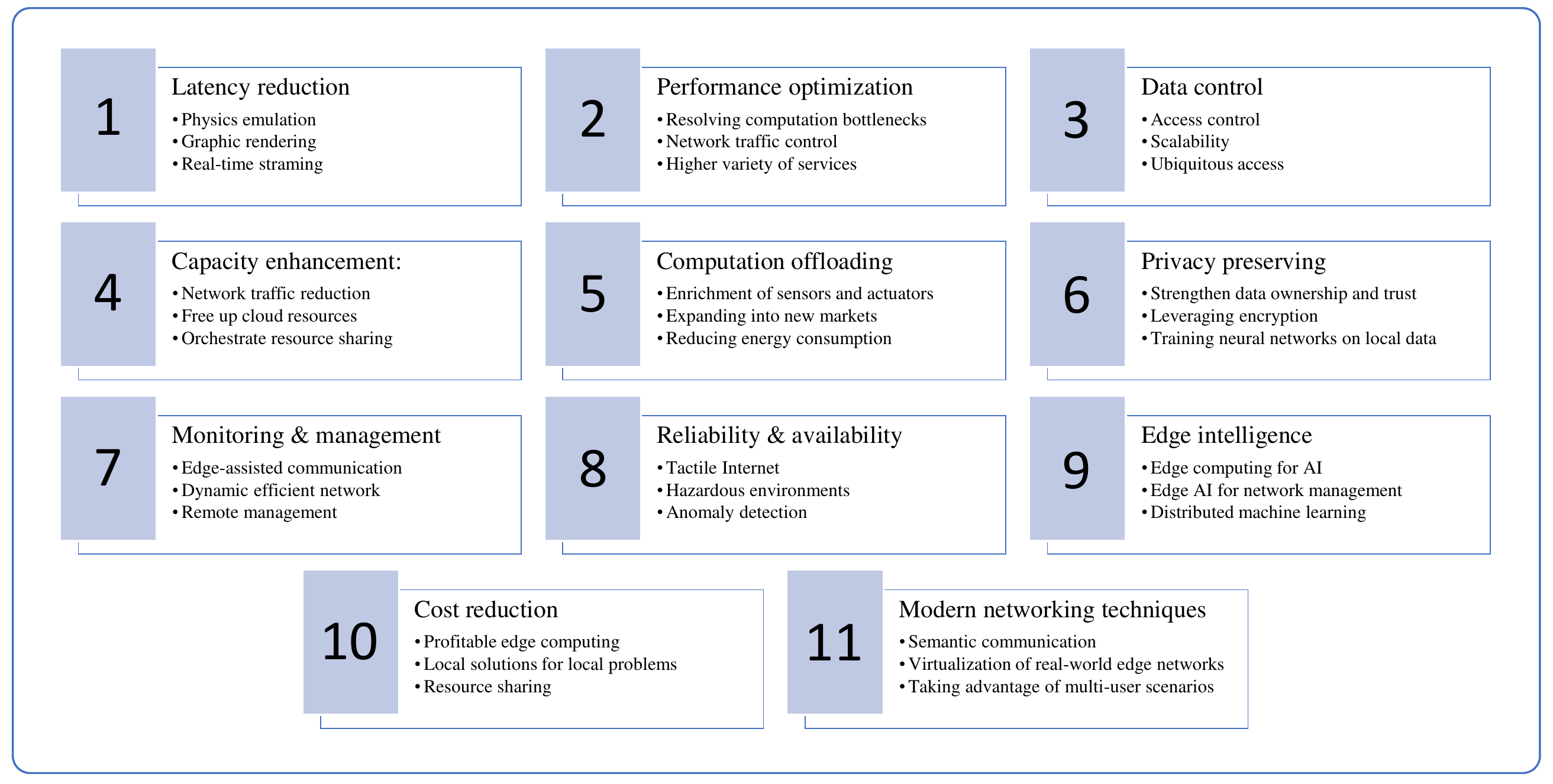}
\caption{Metaverse requirements to be met by edge computing.\label{Edge_figure}}
\end{figure}

\subsection{Metaverse requirements to be met by edge computing}
Many applications in the metaverse need seamless synchronization of the real and virtual worlds. Some of these applications are critical, such as remote control of a robotic assisted surgery system, and some of them, such as guiding an avatar in a virtual environment, are related to the quality of the user experience (QoE). Therefore, it is necessary to provide a metaverse infrastructure to provide \emph{low-latency services}. In traditional solutions, processing and storage are outsourced to cloud servers. These servers can easily be located thousands of kilometers away from users, and this distance causes a lot of delays in telecommunications. Edge computing, as an alternative solution, brings processing and storage closer to users and the place where data is generated. Hence, in the literature, some metaverse processes that are very apt to run on edge servers and edge devices such as physics emulation \cite{kechadi2022edge, lim2022realizing}, graphic rendering \cite{xu2022full, mehrabi2021multi, xie2021sharing, xie2022qos}, and real-time streaming protocols \cite{pan41063155g} have been investigated. 

When edge servers are set up near users, it does not mean that the operator transfers some calculations from the cloud server to the edge server forever. Instead, it flexibly uses edge facilities to optimize network performance depending on different conditions. In other words, edge computing and edge storage capabilities add new dimensions to \emph{network performance optimization problems} and enlarge their feasible region. Therefore, the network has several more degrees of freedom. The operator can dynamically decide, depending on the situation, how the edge servers come into play and bring advantages to the metaverse network. For example, edge servers schedule computing tasks \cite{chen2021recent}, provide better coverage \cite{kamel2019uplink}, run load-balancing algorithms \cite{6GArticle}, resolve computation bottlenecks \cite{mijuskovic2021resource}, and provide heterogeneous services well addressed in the MEC (Multi-access Edge Computing) standard \cite{sabella2019multi}. In \cite{hashash2022towards}, the authors have proposed the idea of using distributed sub-metaverses in such a way that each one is connected to an edge server and covers a part of the space. In this article, they have presented an optimal method for partitioning the space in such a way that the synchronization time of the physical and virtual worlds is minimized. Furthermore, resource allocation considerations between edge servers and the requirements for latency in different parts of the space are taken into account. 

To make the metaverse network \emph{scalable}, it is necessary to support issue tracking, data quality management, data protection, and risk control functions. On the one hand, we know that the metaverse uses distributed systems. It also captures sensitive data such as biometric data, eye movements, and financial data. In addition, the metaverse interacts with a wide range of technologies from blockchain to artificial intelligence. For this reason, the impact of \emph{data control functions} in the metaverse is significant. On the other hand, we know that edge computing enables better and faster data control by keeping data in the local network and close to the user. Edge computing makes it easier for organizations to control data access to a universal metaverse \cite{kechadi2022edge} whilst manages data exchange in various ways such as W-LAN, cellular networks, and satellite communication systems which are supported by 5G and 6G \cite{al2021survey}. It is also possible to support distributed computing and use the processing and storage resources of end-devices, which can help network scalability \cite{liu2019survey}.

Furthermore, small devices may not be able to run many heavy tasks in the metaverse like the inference of large neural networks and real-time renderings. In this case, they should be able to \emph{outsource} these tasks to powerful servers or perform the tasks with the help of other devices in a distributed scheme \cite{jiang2021reliable}. This reduces telecommunications and thus reduces energy consumption \cite{preist2019evaluating}. Edge servers can provide more performance, flexibility, and storage than edge devices. In addition, it is usually easier to upgrade edge servers than end devices. This is because we can equip the edge server with more GPUs, RAM cards, and storage devices, or change the previous components to increase the power of the edge server. But we may not be able to disassemble all edge devices, including IoT modules, and change their features. So far, a lot of research has been done on the outsourcing of computing of IoT devices to edge servers, in which its advantages, challenges, architectures, and various approaches have been explained \cite{elgendy2022survey, lin2020survey}. Further, some IoT devices have limited batteries, and energy costs can be high. In any case, optimizing energy consumption in the Metaverse network is crucial. A practical case study is presented in \cite{mocnej2018impact} that explains the impact of edge computing on energy consumption in IoT devices.

Even with modest Metaverse assumptions, data usage could easily expand more than 20x during this decade \cite{Suissereport}. If edge servers perform storage and processing tasks for users, there is no need to send large amounts of raw data deep into the network. For this reason, the telecommunication capacity of the network is less occupied. Hence, edge computing can replace cloud computing at a lower cost in many applications. Additionally, since there are a lot of powerful edge devices in the metaverse network, it is possible to encourage resource sharing among edge computing applications. Thus, edge devices can serve as microcenters for storage and processing. In this case, the \emph{capacity of the metaverse network} will increase without occupying the resources of the cloud server \cite{petri2017incentivising}. The authors in \cite{liu2018erp} have presented an efficient method for sharing resources at the edge of the network in the framework of the resource pool. For a survey of incentive mechanisms for sharing resources at the network edge, see \cite{huang2022incentive}.

Privacy protection is another challenge in the metaverse. Storing data on a local edge server (for example, at a university, hospital, or airplane) can help keep users' data confidential \cite{ng2021unified}. Edge servers can monitor the passage of sensitive data and be equipped with hardware and software protection tools. Many IoT sensors and devices are not smart enough to not be compromised. Sometimes their communication is not secure enough. Having an edge server near these devices and possibly owned by the user can maintain confidentiality and bring more trust. Users connected to an edge server can use local area network policies such as device and file sharing. On the other hand, authentication mechanisms can be implemented on edge servers to secure the network against various types of hacker attacks \cite{singh2020hierarchical}. Edge computing can come to the aid of end devices and participate in cryptographic algorithm calculations \cite{kumari2022preserving} and training neural networks on local data \cite{pham2022artificial}. 

Edge servers make the network more self-aware and easy to manage and monitor. Spending on creating sites for edge servers can be multi-purpose and simultaneously address several different needs. Edge servers provide monitoring of local network information and its devices and help with higher-level configuration and customization. Edge computing is a key technology in the achievement of 5G goals. It is also one of the enablers of 6G networks \cite{dong2022edge}. It enables distributed scenarios and low-latency telecommunications which can help improve network traffic and be part of upcoming solutions in telecommunication issues such as channel selection. Furthermore, to choose the most appropriate coding and communication protocol, edge servers can send network conditions to end devices and vice versa \cite{cheng2022md2do, chang20226g}. Edge computing tools help end devices in many ways. In addition to performing calculations, storage, caching, prefetching, and data traffic prediction of end devices, they can improve network efficiency and dynamics. For example, they can have different communication channels and direct the information packets of connected devices from the most efficient path \cite{marvi2020toward}. Additionally, solutions have been offered to avoid the need for physical presence to calibrate end devices with the help of edge computing \cite{wang2022remote}. Environmental threats such as temperature rise, humidity, dust, and power outages are just as serious as cyber threats and can put sensitive edge devices out of service \cite{Edge_monotoring}. The presence of processing and storage facilities at edge computing sites enables edge remote management solutions. With the help of these solutions, we can take care of the critical facilities of the Metaverse network at the edge and reduce the energy required for troubleshooting and repair.

Additionally, we want the metaverse network to be more \emph{reliable and available}. Although there is no universal definition of reliability and availability, and there is no standard way to measure them, we can talk about strengthening the reliability and availability of some services compared to others. In some applications that are offered under MaaS, these two features are prominent and should be considered in design, programming, and warranties given in agreements. In the event of a network failure, edge computing infrastructure can still maintain the connection of the devices and perform some processing. Additionally, it is possible to add redundancy to the system at edge computing sites for specific applications. By bringing the processing tasks to the edge servers, the connection of the devices to the core network is reduced. This reduces the level of vulnerability of the system to cyber threats. In \cite{jiang2022reliable}, to increase the reliability of the Metaverse network, the authors have presented a method based on distributed coded computing algorithms that can be implemented at the edge. Also, a growing number of solutions are being offered for edge computing reliability in hazardous environments \cite{Lanternedge}. System reliability is significantly affected by the detection and prediction of anomalies. Artificial intelligence can train machine learning models on the large amount of local data that edge servers have access to and use them in real-time \cite{yu2022edge}.

3D reconstruction, character animation, detection of harmful content, machine vision, gesture recognition, facial expressions, understanding emotions, body movement recognition, physical interactions, eye tracking, brain-computer interface, and many more are activated by artificial intelligence in the Metaverse. Some strategies for using machine learning to strengthen security in the Metaverse and to create recommendation systems are proposed in \cite{ghantous2022empowering}. For various reasons, there is a rising desire to train and execute machine learning models on edge servers, which is called Edge AI or Edge Intelligence. Many applications in the metaverse such as autonomous vehicles, intelligent interaction of avatars with the environment in the virtual world, health care, smart city, and industrial complexes need machine learning algorithms \cite{mendez2022edge}. Having computing servers on the edge equipped with artificial intelligence allows better network and telecommunication management. For example, in situations where we have many sensors in an environment and we don't want their massive data to leave the local network raw, we can reduce the traffic volume with the help of an edge server that can process their data and reduce their size. In addition, in the edge server, algorithms can be implemented to predict traffic and cache popular data, which increases the service quality. Today, machine learning has been used in different telecommunication layers, from the physical layer and coding of telecommunication signals \cite{jamali2022productae} to the detection of cyberattacks in the application layer \cite{almaraz2022transport}. Designing distributed machine learning algorithms such as federated learning \cite{abreha2022federated}, split learning \cite{thapa2022splitfed} and gossip learning \cite{tu2022asynchronous} is a hot research topic because it adds new capabilities to the network. For example, with distributed learning, the neural network is trained without the data leaving the device and the processing load is spread across multiple devices. In addition, in a local area network where Internet traffic is limited, devices can continue to train neural networks with the help of each other without having a permanent connection to the Internet \cite{filho2022systematic}.

As previously mentioned, edge computing is not necessarily a substitute for cloud computing. Rather, it is an option with its advantages and challenges. The \emph{costs} and desired quality of the various services provided in the metaverse should be examined and the appropriate way to provide them should be chosen. In addition, attention should be paid to the interaction of cloud servers and edge servers and how to divide their work. Due to the existence of various end devices and network equipment that can be used in the metaverse, the importance of this review increases. Edge computing reduces network traffic load and reduces costs. Additionally, it may be cheaper for some businesses to build edge servers than to outsource computing. For example, in some regions, the cost of electricity makes it cost-effective to install edge servers. Cost-effective solutions for edge servers are also provided. For example, the study \cite{avelar2017cost} shows that micro data centers at the edge cost 42\% less than traditional centralized data centers.

In telecommunications, there are several techniques to reduce costs and increase capabilities. Some of them are in the common domain of edge computing and the metaverse. Network designers can use these ideas. For example, semantic and goal-oriented communication is provided to increase stability and efficiency \cite{luo2022semantic}, which can be used in the 6G network and contribute to the metaverse. Edge servers can be equipped with various neural networks and perform semantic communication tasks. On the other hand, the metaverse can help itself is by creating digital twins of edge devices. By monitoring the digital twins of edge devices and edge infrastructures such as edge servers, I/O ports, RIS (re-configurable intelligent surface), UAV (unmanned aerial vehicles), SAGIN (space-air-ground integrated network), and network equipment, the MaaS operator can instantly control and calibrate these physical entities. 

\subsection{Edge computing challenges}
An operator who wants to use the edge computing paradigm in MaaS design must consider several challenges. Limitations of edge servers is one of these challenges. Due to the limitations of end devices (power, bandwidth, computing, storage, etc.), edge servers are proposed to assist end devices in tasks such as artificial intelligence processing and high-quality image rendering. But it should be noted that the power of edge servers can also be limited due to hardware limitations or server room space. In this case, the edge server cannot provide all the desired functions for a large number of users at the same time.

Another challenge to be considered is the resource allocation problem. Users in a region will likely make many requests to the edge server at the same time and we will see network spikes. Managing these requests is a challenging issue. Another issue is the management of faulty hardware. Cloud servers have more infrastructure compared to edge servers. The reason why it is beneficial to have spare hardware on hand is so that in the event of a hardware failure, it is possible to replace it quickly. But on edge servers, this issue is not economical. Suppose you own one device, and you buy a spare. The cost will double. But suppose you have 100 pieces of equipment. If you buy a spare, your cost will be 101 instead of 100, which is relatively not much. In addition, fairness is also essential in resource management. Because there is a lot of resource sharing in the services provided at the edge, the metaverse operator at the edge needs to be very careful about the optimal scenarios. For more information on resource management at the edge, see \cite{zhou2021fairness}.

Stragglers and malicious node are serious problems in outsourcing scenarios at the edge. When a computing load is distributed among several servers or data is stored in several servers, it is possible that some of these servers, i.e., stragglers, will take more time than usual to respond. This issue causes our entire work to be delayed on those servers. Malicious nodes in the network are entities that seek to disrupt the functioning of the metaverse. There should be pre-prepared scenarios to deal with stragglers and malicious nodes. There are various solutions to mitigate stragglers and malicious nodes in the network. For example, in coded computing algorithms, we can divide a task into several parts and outsource it to several servers. If some of these servers are stragglers or malignant, the computation's final answer can still be recovered \cite{jahani2022berrut}.

There are also a couple of communication and networking challenges in using edge computing for the metaverse. Metaverse services have diverse and strict requirements. For example, touch Internet requires high-reliability telecommunications, AR/VR headsets require high bandwidth, and IoT devices require high coverage. A large amount of data in the network and its dynamic nature should not cause friction and decrease the efficiency of the system in terms of reliability, delay, and rate. Increasing the variety of services can make MaaS faster and more stable. For example, if different service modes and qualities are available, for example, the video size can be changed or artificial intelligence can be adjusted with different accuracy levels, some of the network challenges in overload conditions will be solved.

The distributed and heterogeneous nature of the metaverse network at the edge complicates its management too. In addition, edge servers must be synchronized with end devices and cloud servers. In addition, various functions must be executed simultaneously to maintain network performance, such as anomaly detection algorithms, resource provisioning, workload prediction, and traffic routing. In places where edge servers are installed, such as a hospital, a network specialist is not always available. For this reason, another challenge of network management at the edge is the high cost of admin intervention for configuration, updates, and equipment maintenance.

Ownership considerations are another critical issue in this matter. Equipment, data, and software on edge servers can have different owners. For example, in the case of equipment, some edge servers are privately owned, some are cooperatively owned, some are rented out to generate revenue, and some are created with specific access requirements. This issue can make it difficult for the Metaverse operator to enter into contracts and launch new services.

As the number of end devices, edge servers, and IoT modules increases, the metaverse network becomes challenged in various sectors like wireless communication, queue management, and user authentication. For example, in wireless communication, interference and congestion can occur. In queue management, there is a possibility of data being thrown away, and user authentication can take a long time. We also know that edge servers hold sensitive data generated in places like hospitals, financial institutions, and homes. But keeping data close to the user is not enough. Users and businesses should be given access and data confidentiality guarantees. For more information on edge privacy solutions, see \cite{ranaweera2021survey}.

In some applications, it is necessary to know who is responsible for each operation at the edge of the network. It should also be possible to receive complaints and prevent manipulation, corruption, and abuse. Decision-making processes and functions should be expanded to the extent that the privacy and property rights of individuals are not compromised. The participation of different people and communities in monitoring the network not only improves the performance of edge servers in one place but also creates knowledge that can be used throughout the metaverse network.

The needs of users are not the same in all regions. In addition, the data formats available in the network and devices connected to the network are diverse. Therefore, the MaaS operator must design and install edge server equipment and software in a flexible and compatible structure and configuration. Another issue is changes and developments in the network. The network can change over time in terms of users or cloud servers. Accordingly, edge equipment should be able to meet these requirements at a low cost.

\section{Blockchain Technology}
To provide users with the best experiences possible, the metaverse gathers enormous amounts of private data. This information is required by the businesses or programs in order to construct targeted systems successfully. With its authentication, access control, and consensus methods, blockchain gives consumers total control over their data, protecting their personal information \cite{guo2022survey}. 

The blockchain utilizes hash algorithms and asymmetric-key encryption to protect data in the metaverse. Additionally, the quality of the real-world data that users exchange is crucial to the construction of objects in the metaverse. This data is collected from numerous applications, including those in entertainment, healthcare, and more. Blockchain enables individuals and businesses to validate all transactions by providing comprehensive audit trails of all transactions. The metaverse's data quality will improve as a result \cite{gadekallu2022blockchain}.

\begin{figure}[h]
\includegraphics[width=\textwidth]{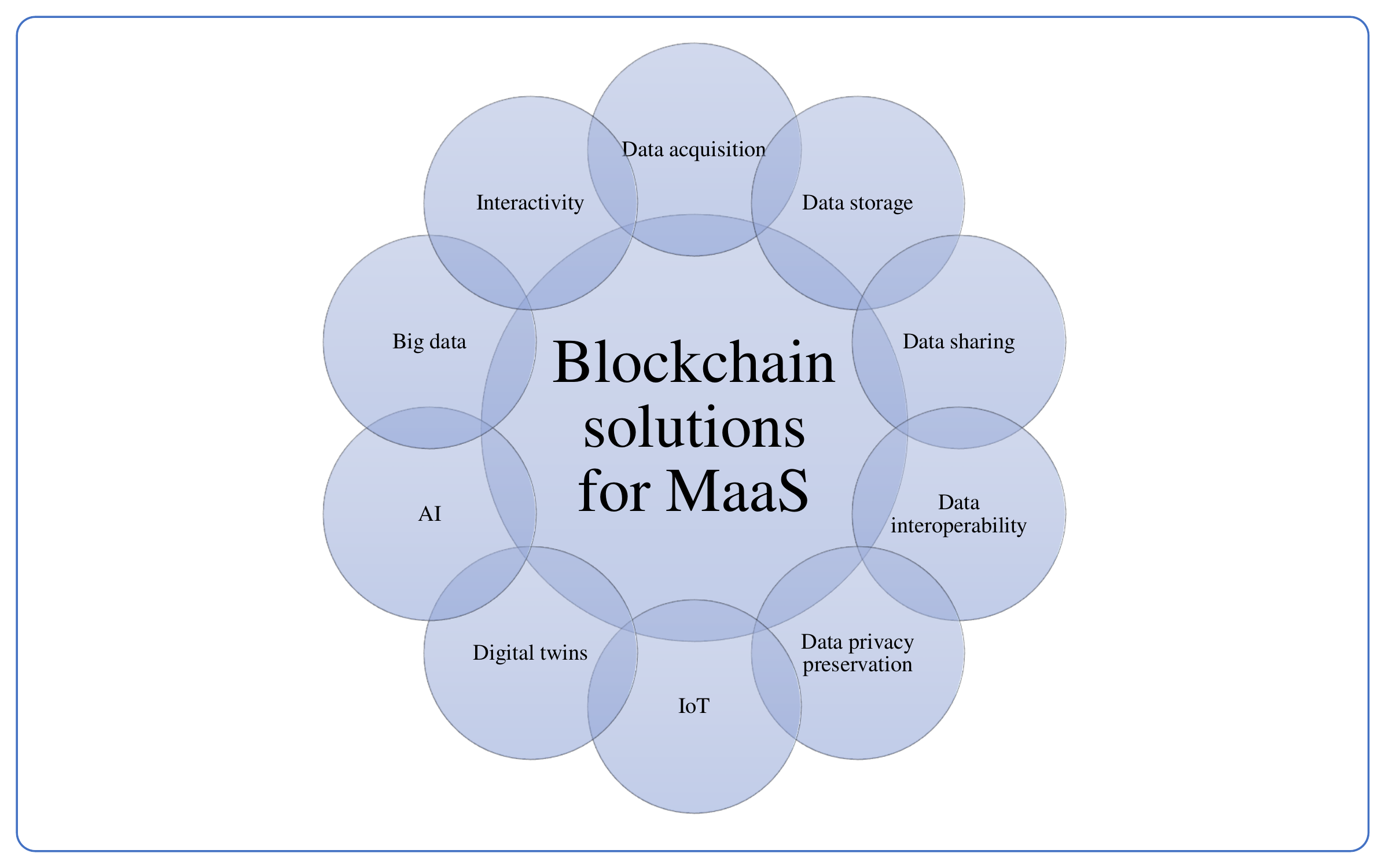}
\caption{Blockchain solutions for MaaS.\label{Blockchain_figure}}
\end{figure}

On the other hand, the successful exchange of AR and VR data is essential to Metaverse. The smooth and safe data sharing within the metaverse is made possible by the blockchain's cutting-edge encoding information system. Moreover, stakeholders in the metaverse must have access to and control over resources in various virtual environments. Due to the many settings in which these virtual worlds are created, data interchange is constrained. A cross-chain protocol enables data interchange between two or more blockchains that are present in different virtual worlds. In terms of integrity, data in the metaverse must be constantly and accurately updated. Due to the immutability offered by the blockchain, the metaverse data is kept as a copy in each block along the chain and cannot be changed or withdrawn without the agreement of a majority of the participants.

A number of vendors have already begun to enter the MaaS market; according to an announcement made by Propel, a blockchain solutions platform, it would provide MaaS solutions for smart contracts, NFT utilities, and decentralized finance (DeFi). Moreover, Lovelace, a different blockchain-based platform, offers a MaaS toolkit that gives users and developers the technology required to build and trade NFTs, run smart contracts, monetize VR gaming, interact with other metaverse platforms, and much more \cite{calzada2021data}. In the following, the authors concentrate on the steps that a MaaS developer should do and utilize blockchain technology to address various difficulties associated with creating and developing the Metaverse platform. See Figure \ref{Blockchain_figure}.

\subsection{Open challenges for MaaS}
\textbf{Data Acquisition}: The metaverse will generate large amounts of unstructured, real-time data through decentralized applications, but acquiring this data can be difficult. Building applications in the metaverse, such as recommender systems, will require high levels of data integrity. The use of virtual reality and increased streaming in the metaverse will further strain data acquisition systems \cite{brunschwig2021towards, jeong2022rethinking}. The quality of the data may also be impacted by the acquisition of duplicative or incorrect data \cite{shiau2022scale}.

\textbf{Data Storage}: Once the metaverse is fully functional, the physical world's ability to store data may be strained to its breaking point. This could create significant challenges for the metaverse \cite{bian2021demystifying}. If the metaverse relies on a central storage system, there is a risk of data leakage, manipulation, or loss. The potential of the metaverse to offer biometric data, voice inflections, and vital signs that depend on sensitive data is also jeopardized by the likelihood of data loss and corruption in centralized applications \cite{wang2021joint, kiong2022metaverse}. 

\textbf{Data Sharing}: Data sharing on centralized platforms carries the risk of exposing private and sensitive information \cite{liu2020blockchain, egliston2021critical}. Additionally, due to data mutability, there is a risk of high latency and reduced data availability \cite{yu2021blockchain}. This is particularly relevant in the metaverse, where many applications will generate large amounts of real-time data. When demand for real-time data increases, data flexibility can become a problem.

\textbf{Data Interoperability}: The metaverse will be created through the merger of many digital domains, but these domains are currently fragmented and disorganized. This can make it difficult for users to engage with multiple virtual worlds, as they must set up separate accounts, avatars, hardware, and payment infrastructure for each one \cite{bian2021demystifying}. There are also few methods for users to transfer their digital assets between different digital environments. In order for the metaverse to be truly interoperable, digital world apps must be able to easily exchange information with one another, regardless of their location or the technology being used. The conventional approach to interoperability is inadequate for the metaverse, so new solutions are needed \cite{mystakidis2022metaverse}.

\textbf{Data Privacy Preservation}: In the early stages of the metaverse, attackers may be able to deceive users and steal important data. This could be particularly dangerous if attackers use artificial intelligence bots, as users may not be aware that they are not speaking with a real person. The metaverse also raises concerns about the confidentiality of personal data, particularly personally identifiable information (PII) \cite{hughes2022metaverse}. Finally, as the metaverse grows and more validity information is included, managing the large amounts of data will become increasingly difficult.

\textbf{IoT}: The metaverse will have a large number of interconnected IoT sensors, which raises concerns about IoT security and storage. Real-time analysis of unstructured IoT data is also challenging \cite{zhang2021blockchain}. When storing data across virtual worlds, a centralized solution is not ideal, as tampering with even one piece of data could compromise the entire set of findings. Additionally, data sharing across virtual worlds will depend on the cross-platform capabilities of IoT devices \cite{hajjaji2021big}. Finally, IoT data tracking is necessary for safety and legal compliance.

\textbf{Digital Twins}: The quality of the data used to build digital twin models is important for their accuracy, so the information provided by the source must be accurate and of high quality \cite{zhuang2021digital}. Digital twins from different sectors, such as healthcare and finance, must be able to communicate and connect with one another. To improve the accuracy and consistency of communication, digital twins should be able to detect and correct faults. However, data security can be a challenge when using a range of devices and sensors to create digital twin models that utilize real-time data. This can be particularly vulnerable to botnets and other viruses \cite{khan2022digital}.

\textbf{AI}: The ownership of AI-powered content in the metaverse is difficult to determine, as users have no way to distinguish between communicating with a real person and an avatar created by a computer. This could lead to users using AI technology to exploit other users or resources in the metaverse, such as by cheating at games or stealing from other users' accounts \cite{wiederhold2022ready}. Additionally, AI may make mistakes, which could lead to people losing trust in the metaverse. Another challenge is the use of a similar blockchain across different AI applications in the metaverse.

\textbf{Big Data}: One of the main challenges is the sheer volume and rate of data production in the metaverse, which can be difficult to keep up with, even with advances in data storage technology. Another challenge is the variety of data produced by metaverse apps, which can make it difficult and time-consuming to collect and organize the necessary data for consumers. Additionally, the rapid development of big data technology \cite{ahsani2020improving, ahsani2019quantitative, sharma2019assessing, chakraborty2018traffic, poddar2018comparison, sharma2017evaluation} can make it challenging to stay up to date with technical advancements in the metaverse. 

\textbf{Interactivity}: The metaverse, which is a virtual world created through the use of technology like holographic telepresence and augmented reality, offers immersive, realistic experiences by combining audio, video, cognition, and other elements. However, the use of XR technology in the metaverse also creates challenges related to data storage, data sharing, and data interoperability. For example, the businesses can create recommendation systems using data from XR technology, but this data must be stored securely and shared transparently with stakeholders. Additionally, the metaverse must be able to handle the exchange of data between virtual worlds in an interoperable way in order to provide users with a seamless experience \cite{bhattacharya2021coalition}.

\subsection{Blockchain solutions for MaaS}
\textbf{Data Acquisition}: With the use of blockchain technology, it will be simpler to gather reliable data in the metaverse for uses like social networking. Blockchain's distributed ledger will make it possible to trace data in the metaverse and validate transaction records \cite{islam2019buav, deepa2022survey}. Because the majority of nodes in the ledger must consent before any modifications to the data in the metaverse can be made, data collection is therefore resistant to attacks \cite{xu2021lightweight}. A blockchain-specific validation process that is driven by consensus mechanisms is applied to all data collected in the metaverse \cite{bouraga2021taxonomy, lashkari2021comprehensive}. Every action is documented as a transaction on a blockchain, and each block includes a cryptographic hash of the one before it, as well as the metadata, a date, and the activity \cite{luo2021novel}. As a result, changing the data in one block will change the data in all the other blocks as well. Any block's data is impervious to manipulation \cite{zhang2021research}. There will be no repetition in the data collecting process since the likelihood of producing a duplicate block is almost zero. Data obtained by blockchain enabled acquisition mechanisms in the metaverse will be trustworthy since every block is approved on the blockchain \cite{guo2021reliable}.

\textbf{Data Storage}: The metaverse storage is impermeable to hacking since a new block is generated for each transaction \cite{liang2020secure}. As a result, data is stored across the chain as a copy of the original blocks, increasing data dependability and transparency in the metaverse \cite{jeon2022blockchain}. If the centralized data store is hacked, the metaverse applications, which include anything from real estate to digital things, would be very vulnerable \cite{yang2022expert}. Utilizing blockchain technology will lead to a large number of blocks contributing to data distribution, enhancing data accessibility in applications like vital monitoring and life support alerts in the metaverse. Blockchain technology's decentralized nature enables data scientists in the metaverse to work together and on data cleaning, which will greatly minimize the time and expenses involved with labeling data and getting datasets ready for analytics \cite{Ahsani42}.

\textbf{Data Sharing}: Blockchain technology has the potential to increase the accuracy and transparency of transactions in the metaverse for applications such as education and cryptocurrency trading \cite{egliston2021critical}. Stakeholders would be able to access a decentralized, unchangeable record of all transactions created by applications like governance and finance. Therefore, increased data openness will be advantageous to the metaverse's stakeholders \cite{rashid2021blockchain}. Users' confidence will increase as a result of being able to comprehend how third-party programs like Thunderbird, the Bat, and Pegasus manage data thanks to blockchain technology, which can also reduce grey market transactions \cite{vashistha2021echain}. The owner of the data will also have total control over the data. Distributed ledger technology can also be useful for data audits. Blockchain thus saves time and money by reducing the need for data validation \cite{min2022portrait}. The flexibility of data sharing will be increased through smart contracts. Usually, they are used to automate the execution of a contract so that all parties may be sure of the result right away, without the need for an intermediary or a waste of time. The diverse programming of smart contracts is made possible by blockchain. As a result, programs like Nmusik, Ascribe, Tracr, UBS, and Applicature will benefit \cite{ali2021comparative}.

\textbf{Data Interoperability}: A cross-chain protocol is the ideal approach to guarantee interoperability across virtual worlds in the metaverse \cite{belchior2021survey, madine2021appxchain}. This enables the transfer of goods like avatars, NFTs, and money between virtual worlds. This protocol will lay the foundation for broad acceptance of the metaverse. Cross-blockchain technology will make it possible for virtual worlds to communicate with one another, doing away with the necessity for middlemen in the metaverse \cite{jabbar2020blockchain}. In the metaverse, connecting users and apps will be made simple via blockchain.

\textbf{Data Privacy Preservation}: Through the use of private and public keys, blockchain technology enables users of the metaverse to govern their data, effectively giving them ownership over it. Third-party intermediates are prohibited from misusing or obtaining data from other parties in the blockchain-enabled metaverse. Owners of personal data stored in the blockchain-enabled metaverse will have control over when and how third parties can access that data \cite{kumar2021ppsf}. Blockchain ledgers come with an audit trail as a standard, guaranteeing that the transactions in the metaverse are comprehensive and consistent. Zero-knowledge proof has been used on the blockchain, giving people easy access to the identification of crucial data in the metaverse while keeping their privacy and control over their belongings. Blockchain technology uses zero-knowledge proofs as a method for users to convince apps of something without having to provide the information \cite{sedlmeir2021next}.

\textbf{IoT}: Through cross-chain networks, which are created by blockchain technology, IoT devices in the metaverse may exchange data and create tamper-proof records of shared transactions in virtual worlds. Applications and individuals will be able to exchange and access IoT data thanks to blockchain technology without the requirement for centralized administration or control \cite{majeed2021blockchain}. Each transaction is documented and validated in order to reduce disputes and boost user confidence throughout the metaverse. IoT-enabled blockchain in the metaverse makes it possible to store data in real-time. Due to the immutability of blockchain transactions, all stakeholders can depend on the information and respond quickly and effectively \cite{dorri2021temporary}. By allowing stakeholders to manage their IoT data records in shared blockchain ledgers, blockchain technology can assist in resolving problems in the metaverse.

\textbf{Digital Twins}: Digital twins are attack-resistant because to blockchain's encryption capabilities and historical data openness, which also allow for safe data sharing \cite{lee2021integrated} across many virtual worlds. With the use of an intelligent distributed ledger, data may be exchanged between digital twins in virtual environments. Using an intelligent distributed ledger, real-world items will be saved on the blockchain and synced to digital twins in the metaverse. The implementation of digital twins on a blockchain will also help to resolve problems with data security and privacy \cite{shen2021secure}. Tracking sensor data and creating high-caliber digital twins in the metaverse will be possible by combining blockchain and AI. Every digital twin activity in the metaverse will be documented as a transaction on the blockchain, which is unchangeable and requires consensus to modify \cite{lee2021integrated}.

\textbf{AI}: Blockchain-based encryption gives users of the metaverse total control over their data and makes it easy to transfer ownership of AI consent to another entity. Through the use of zero-knowledge proofs, users may convince apps and other parties that certain information about them is true without divulging this information to the applications themselves, granting the authority to utilize data for AI model training. Blockchain ledgers frequently offer an audit trail that may be used to verify the legitimacy of any transactions that take place in the metaverse. People can locate crucial metaverse facts via a zero-knowledge evidence system while still maintaining their privacy and control of their resources against deepfakes \cite{hussain2021artificial}. By doing this, AI will be stopped from wasting resources in the metaverse.

\textbf{Big Data}: By assisting in the collecting of data from reliable data sources, blockchain technology will help to reduce the quantity of inaccurate data received.  Data modification by other parties will be prohibited, and the data owners will have complete control over their data. This guarantees high-quality data flows across the metaverse \cite{deepa2022survey}. Data scientists in the metaverse will be able to communicate and work together on data cleaning thanks to the decentralized nature of blockchain technology. This will greatly cut down on the time and costs involved in categorizing data and building datasets for analytics applications, as well as the danger of data contamination. Since the data will be copied across the network and the blockchain is immutable, it will be impossible to alter it \cite{gligor2021theorizing}. Data accessibility for metaverse stakeholders will therefore be improved.

\textbf{Interactivity}: A blockchain-based distributed ledger would make it possible to validate the records of holographic telepresence and other XR applications in the metaverse and track the origin of inaccurate data. As a result, a more precise recommendation system will be created. The zero-trust mechanism and cross-chain technology of the blockchain will make it simpler for holographic telepresence and other XR applications to safely transmit data between virtual worlds \cite{bhattacharya2021coalition}. Data integrity is guaranteed for XR apps and holographic telepresence by the interplanetary file system offered by the blockchain. The consensus technique used by these devices will make the data they gather and store on a blockchain unchangeable. Blockchain supports confidence among AR/VR stakeholders by facilitating transparent ownership transfer and asset verification \cite{kumar2021decentralized}.

\section{Future Visions and Directions}
\subsection{Content-centric Metaverse}
With the ever-growing increase in using MaaS platforms, UGC is expected to be increasingly generated and transmitted through the metaverse networks. Current IP-based host oriented content transmission protocols will face critical challenges for securing UGC dissemination by heterogeneous end devices over the metaverse platforms. To address this challenge, content centric networking (CCN) can be employed to rethink the current Internet architecture. According to CCN, contents will be routed directly by their naming information instead of IP addresses \cite{tourani2017security}. For data and content sharing in CCN-based metaverse, users request the desired UGC via sending an interest message to any CCN based node that occupies the matched content. The main idea for securing CCN is to directly safeguard the security of every single content/data itself, rather than securing the “pipe” or the communication channels/links \cite{ghali2016futility}. In this way, flexible and content-centric secure metaverse can be realized. Due to the inherent attributes of the CCN architectures, CCN-based metaverse can explicitly cause new security concerns as well. For instance, content poisoning and network monitoring would be two security issues in the CCN-based metaverse. Specifically, malicious users might inject poisoned UGCs, resulting in delayed or failed valid UGC delivery, e.g., through flooding. A curious CCN node might observe the sensitive content disseminated by CCN users by directly monitoring the network traffic. Therefore, further research on privacy and security protection for CCN-based metaverse is required \cite{yang2020sgx}.

\subsection{Edge computing}
Hospitals, production lines, residential complexes, offshore equipment and game centers have different requirements. Hence, solutions based on edge computing in the Metaverse should be adapted for different applications. This work requires a methodical, repeatable and well-reasoned approach from the Metaverse operators. Otherwise, the cost of maintaining and managing edge servers will increase and many problems will arise for users and the network. At the same time, the variety of services and flexibility in service level agreements should still exist so that users have a better user experience.

For various applications, edge computing can be used to increase reliability, reduce latency, and offload tasks in 5G networks \cite{coutinho2022design}. More interestingly, edge computing is one of the main enablers of the 6G network because it can be used for reliable low-latency communications, AI-empowered capabilities, and increasing energy efficiency \cite{bandi2022review}. In addition, edge computing plays a vital role in the new generations of the Internet of Things \cite{Europe_iot}. Hence, we expect edge computing to progress along with related technologies and add new capabilities to the metaverse.

In designing edge computing solutions, various tradeoffs must be considered. Processing power, storage volume, telecommunication link bandwidth, spare parts, emergency power, security systems and many other things must be selected depending on the needs of users. In addition, it is possible that the needs of users change over time. A scalable and flexible design approach can help with this. In this type of design, it is possible to increase and decrease each of these facilities, depending on the conditions. For example, it is not necessary for the Metaverse operator to have installed a lot of storage space on the edge sites from the beginning. However, it should have provided the possibility of increasing storage space on the edge servers. This would enable the network to meet this critical requirement in the shortest possible time with the increase in users' needs. 

\subsection{Blockchain’s still unsolved challenges}
Despite many challenges which were discussed in this paper for MaaS developers and the solutions that blockchain technology can provide, there are still more constraints that should be taken into consideration. There is still work to be done on creating a robust blockchain for the metaverse to overcome unsolved problems. In the following, the authors mention some other unsolved challenges and make some recommendations as well.
\begin{itemize}
\item Blockchain can be slow because of its complexity and distributed nature.
\item On a blockchain, transactions can take a very long time to execute.
\item Data in the metaverse will be more able to withstand copying and manipulation with the help of a consensus-based distributed ledger, but since any new data must be duplicated throughout the entire chain, more study is needed to overcome the latency problem.
\item The number of blocks must grow along with the number of users in the metaverse, requiring the employment of enormous computational resources \cite{gao2021b}. As a result, users will pay a higher transaction cost for the verification of shared transactions. For effective data sharing in the metaverse, next-generation blockchains must overcome this problem \cite{rahimi2021multi}.
\item The existence of numerous public blockchains in various virtual reality environments that do not speak the same language presents the biggest obstacle to cross-blockchain enabled the metaverse interoperability. It will be challenging to adjust because different platforms will offer different degrees of smart contract capabilities. Furthermore, these virtual worlds use a wide range of transaction architectures and consensus mechanisms, which limits interoperability \cite{wibowo2019improving}.
\item If a small number of miners control the majority of the network's overall mining hash rate, blockchains are susceptible.
\item Due to the anonymity offered by blockchain technology, it is challenging to track down all IoT transactions involving illicit services in the metaverse.
\item To carry out the metaverse's expansion, the blockchain needs to be regularized.
\item For blockchain to be successfully used in digital twin applications in the metaverse, challenges like standardization, privacy, and scalability must all be resolved.
\item The quality of digital twins in the metaverse will increase as a result of the integration of blockchain, XAI, and federated learning methodologies.
\end{itemize}

\section{Conclusions}
This article provided an overview on privacy and security aspects of the metaverse, from
different perspectives, including the wireless access, learning algorithms, data access, and human-centric interactions. New directions towards realizing privacy-aware and secure metaverse-as-a-service (MaaS) platforms were addressed, and less-investigated methods were reviewed to help mobile network operators and service providers facilitate the realization of secure and private MaaS through different layers of the metaverse, ranging from the access layer to privatizing the social interactions among clients. Additionally, edge computing, which is one of the key enablers of the metaverse, has been discussed, along with the advantages and challenges associated with its use in the metaverse. Later in this work, a comprehensive investigation and analyses of challenges for MaaS developers and the blockchain’s solutions for MaaS platforms were provided. At the final, future vision, unsolved challenges, and some recommendations were also discussed to bring further insights for the network operators and engineers in the era of metaverse.

\bibliographystyle{ieeetr}
\bibliography{references}

\vspace{12pt}

\end{document}